\documentclass[fleqn,usenatbib]{mnras}

\usepackage{newtxtext,newtxmath}
\usepackage[T1]{fontenc}
\usepackage{ae,aecompl}

\usepackage{graphicx}	% Including figure files
\usepackage{amsmath}	% Advanced maths commands
\usepackage{amssymb}	% Extra maths symbols

\title[The Beam Balance]{The Beam Balance - Measuring Binary Systems via Relativistic Beaming Signals from Stars \textit{and} their Companions}

\author[Z. Penoyre]{
Z. Penoyre$^{1}$\thanks{E-mail: zpenoyre@gmail.com}
\\
$^{1}$Institute of Astronomy, Madingley Road, Cambridge CB3 0HA, UK
}

\date{Accepted by MNRAS}

\pubyear{2019}

\begin{document}
\label{firstpage}
\pagerange{\pageref{firstpage}--\pageref{lastpage}}
\maketitle

\begin{abstract}
In this paper I show that the concept of relativistic beaming  - the process by which light emitted by a fast moving sources is lensed towards the direction of motion - can be easily extended to model the signal from both the star and any secondary companions. 

Most companions will be cooler and less massive than their host star. Their lower mass leads to faster orbital velocities, and thus a potentially larger beaming effect. The lower temperature will mean that most of their light is emitted at longer wavelengths, where the relative photometric dominance of the primary is reduced. 

Thus for some systems, the secondary companion can be the main contributor to observed relativistic beaming signals at long wavelengths. Furthermore, if the system is observed over a range of wavelengths we can independently constrain the temperature of the companion, and the mass and radius ratio of the binary. 

To conclude I discuss the current and future observational prospects of this signal, using the properties of known exoplanets to show that such a signal may be observable by upcoming surveys. %Particularly I highlight with telescopes capable of comparable precision to current exoplanet surveys, but observing at slightly longer wavelengths, would likely be able to observe this effect.
\end{abstract}


%%%%%%%%%%%%%%%%%%%%%%%%%%%%%%%%%%%%%%%%%%%%%%%%%%

%%%%%%%%%%%%%%%%% BODY OF PAPER %%%%%%%%%%%%%%%%%%

\section{Introduction}

Relativistic beaming, or simply beaming as I will call it hereafter, is a special-relativistic effect where light emitted isotropically in a rest frame appears - when transformed to a moving frame - to be aberrated. Specifically sources will appear more luminous if moving towards the observer, and dimmer when moving away (see \citealt{Rybicki86}). It is general and ubiquitous - the special relativistic extension of the doppler effect - and it will occur when any body moves and emits light whether it is a hyper-velocity star or a sprinter on the athletics track.

\citet{Loeb03} introduced this concept as a method for finding extra-solar planets and non-luminous stellar companions. The orbital velocity of the star around the barycenter of the star-planet system can be sufficient to cause detectable beaming signatures. This effect has been observed in a small number of systems e.g. \citet{Mazeh12}. 

The modulation of the observed light curve is of order $\frac{v}{c}$, where $v$ is the line of sight velocity of the object and $c$ is the speed of light. At present there are only a handful of planetary systems where the signal will be detectable (\citealt{Loeb03} and \citealt{Penoyre18a}), with a relative change in flux of over 1 ppm (part per million), corresponding to a star with orbital velocities of order 1 kms$^{-1}$. For only a few of these systems will beaming be the dominant source of flux variation, with planet-raised tides, thermal emissions and reflections dominating the out-of-transit lightcurve (see \citet{Penoyre18b} figure 5 which shows which effects dominate for different system parameters). 

The number of known planetary systems with detectable beaming effects may only be in the tens, but with more precise current and future telescopes such as TESS \citep{Ricker15} (which will observe a population of nearer stars than Kepler with correspondingly more precision), JWST \citep{Beichman14} and PLATO \citep{Rauer14} this number may rise dramatically as the achievable precision of these surveys drops from 100s to 10s of ppm per orbit. Of particular interest is the fact that the amplitude of the beaming signal is only weakly affected by distance, so could still be significant for Jupiter mass planets ($\sim$ 1 ppm) at orbits approaching 1 AU, a distance at which the thermal and tidal signal are much reduced \citep{Penoyre18b}.

In this paper I move the focus away from the primary source of luminosity in the system, to its dimmer secondary companion. I shall hereafter just call brighter source the star (though it could be any luminous source) and I'll refer to the less luminous source as the companion. The companion could be a second, smaller and dimmer star, a brown dwarf, a compact object or a planet. In general I make the weak assumption that the companion has lower luminosity, mass and temperature than the star (though I will also discuss more general cases).

Lower luminosity simply means that we can't easily detect the companion (otherwise simpler methods can be used to detect and characterize it). Lower temperature means that its spectrum is significantly different from that of the star (this is also true for higher temperatures and I will discuss this also, but a cooler companion is the more usual case). Finally lower mass means that it will move significantly faster, its speed increased by a factor of $\sim \frac{M}{M_c}$ where $M$ is the mass of the star and $M_c$ that of the companion.

The relative strength of the beaming signal, which will be proportional the ratio of luminosities modulated by the ratio of velocities, can favour the companion. Expressing this symbolically, let the magnitude of the beaming signal of the star be $\delta$, and the flux of the star at some wavelength be $F(\lambda)$. Similarly let the companion have beaming signal and flux $\delta_c$ and $F_c(\lambda)$ respectively. Then the relative magnitude of the beaming signals is
\begin{equation}
\frac{\delta_c}{\delta} \sim \frac{M}{M_c} \frac{F_c(\lambda)}{F(\lambda)}.
\end{equation}
For most systems of interest the first fraction on the right-hand side is large and the second small, but dependent on the system in question their product may be $>1$ at some $\lambda$.

In this paper I will derive the equations governing the modulation of light due to relativistic beaming of a star and its companion (section \ref{theory}) and discuss the present and future possibilities of using such a method and what it can tell us about a system (section \ref{observation}).

\section{Theory}
\label{theory}

\subsection{The basics of relativistic beaming}

%The light we see from a star is some fraction of the total energy it is emitting. The total flux $F$ (energy per unit time emitted) is spread inhomogeneously over a range of wavelengths which we shall denote as
%\begin{equation}
%F_\lambda = \fra{dF}{d\lambda}
%\end{equation}
%such that the total flux emitted between $\lambda$ and $\lambda + d\lambda$ tends to $F_\lambda d\lambda$ as $d\lambda \rightarrow 0$. This can also be thought of as a measure of the number per unit wavelength of photons with some specific energy (and thus wavelength) multiplied by the energy each carries.

%Whether through the naked eye, a telescope or a microwave meal we "observe" the emission from an object via the total energy transferred from photons absorbed by our detector. Most detectors are tuned to recognise a specific range of wavelength (such as the cones in the eye), i.e. are chemically or optically tuned to absorb only photons of a specific energy.
%GOT PART WAY THROUGH WRITING A TEXTBOOK INTRODUCTION TO OBSERVATIONAL ASTRO...

I denote apparent flux observed from an emitting object as
\begin{equation}
F = \int_0^\infty F_\lambda(\lambda) \phi(\lambda) d\lambda.
\end{equation}
$F_\lambda = \frac{d}{d\lambda}F_{emit}$ the amount of energy per unit wavelength emitted by photons with wavelength between $\lambda$ and $\lambda + d\lambda$.

$\phi(\lambda)$ is the \textit{window function} (sometimes called the response function) which tells us what fraction of the energy at a given wavelength is absorbed by our detector, normalised such that $\int_0^{\infty} \phi(\lambda) d\lambda = 1$.

The total beaming effect is the sum of two individual parts:
\begin{itemize}
\item Relativistic beaming of the light - A direct consequence of special relativity which changes $F_\lambda$. The transformation of the momentum and energy of photons emitted isotropically in some rest frame (e.g. that of a star) leads to a lensing of the light along the direction of motion in another frame (i.e. that of a galaxy the star is orbiting) as well as a boost to the energy of photons due to their frequency change.
\item Shifting of the spectrum - The light also experiences a Doppler shift, a purely classical effect, altering the wavelength of the photons in a frame moving relative to the source. Whilst this can be thought of as another change to the spectrum it is elegant to consider it as a shifting of the window function $\phi(\lambda)$.
\end{itemize}

Thus a moving star will have some new apparent flux
\begin{equation}
F'(\lambda) = \int_0^\infty F_\lambda'(\lambda) \phi'(\lambda) d\lambda.
\end{equation}

In this paper I will only consider relatively small changes to the apparent flux caused by motion at speeds much less than the speed of light ($\frac{v}{c} \ll 1$ always) and thus it will be useful to work in terms of the small relative change:
\begin{equation}
\delta = \frac{F' - F}{F} =\frac{\delta F}{F}.
\end{equation}

For subluminal motion we can define, to first order, the effects of relativistic beaming:
\begin{equation}
F_\lambda' = F_\lambda \left( 1 - 4\frac{v}{c} \right) + O(2)
\end{equation}
(see \citealt{Rybicki86} equation 4.97b in the limit $v \ll c$) and of the Doppler effect:
\begin{equation}
\phi'(\lambda) = \phi \left(\lambda(1+ \frac{v}{c} + O(2)) \right) = \phi(\lambda) + \frac{v}{c}\frac{d\phi}{d\lambda} + O(2).
\end{equation}

Thus retaining terms only up to first order
\begin{equation}
\delta = \frac{v}{c}\left(\frac{\int_0^\infty \lambda \frac{d\phi}{d\lambda} F_{\lambda} d\lambda}{\int_0^\infty \phi F_{\lambda} d\lambda}-4\right) + O(2).
\end{equation}

In many cases the spectral flux ($F_\lambda$) may be more easily defined than the window function and it will be convenient to re-express this via an integration by parts as
\begin{equation}
\delta = - \frac{v}{c}\left(5 + \frac{\int_0^\infty \phi \lambda \frac{dF_{\lambda}}{d\lambda} d\lambda}{\int_0^\infty \phi F_{\lambda} d\lambda} \right) + O(2)
\end{equation}
(where I have used the assumption that $\phi(\lambda) \rightarrow 0$ as $\lambda \rightarrow 0, \infty$).

In general the spectrum of the emitting object is some function of wavelength and effective temperature $T$ (other factors like chemical composition or density profile of the atmosphere could be factored in but I will ignore those here). Thus, as I am integrating over all wavelengths, I can express this as
\begin{equation}
\delta = -\beta(T,\phi)\frac{v}{c} + O(2),
\label{beta}
\end{equation}
where $\beta$ is a factor which only depends on an objects motion ($v$), temperature ($T$) and the filter used to observe it ($\phi$).

Everything I have derived until now is completely general, but to express $\beta$ more explicitly I must make some assumption about the spectrum and filter used.

\subsection{Defining the spectrum}

\subsubsection{Power law spectra}

If I assume the spectrum follows a power law in $\lambda$:
\begin{equation}
\label{fApprox}
F_{\lambda} \approx k {\lambda}^\alpha.
\end{equation}
then trivially $\lambda \frac{dF_{\lambda}}{d\lambda} = \alpha F_{\lambda}$ and $\beta = 5+\alpha$.

Though we would not expect any realistic system to obey such a law across the entire spectrum for a sufficiently narrow filter it will be approximately true and thus is a useful estimation. For example, in the Raleigh-Jeans limit of a blackbody spectra $\alpha \rightarrow -4$.

\subsubsection{Blackbody spectrum}

Most spherical astrophysical bodies, across the majority of their spectrum, are well represented by a single temperature blackbody. Though there may be many individual spectral features, when viewed over a finite width window the significance of these features is reduced, meaning that a blackbody spectrum is a passable approximation for objects ranging from giant stars and compact objects to the Earth, the Sun and Jupiter.

A single temperature blackbody follows
\begin{equation}
\label{blackbody}
F_{\lambda} = 2 \pi \left(\frac{R}{D} \right)^2 \frac{h c^2}{\lambda^5} \frac{1}{e^{\frac{hc}{k \lambda T}} - 1},
\end{equation}
where $R$ is the radius of the object, $T$ its temperature and $D$ the distance from Earth ($h$, $k$ and $c$ are Planck's constant, Boltzmanns constant and the speed of light respectively).

It will be useful to define a characteristic wavelength,
\begin{equation}
\label{Lambda}
\Lambda(T) = \frac{h c}{k T},
\end{equation}
roughly where the spectrum peaks. We can summarise all the temperature dependence of a single object in this parameter.

In these terms
\begin{equation}
\label{fSimp}
F = \frac{2 \pi h c^2}{D^2} R^2 \mu(\Lambda,\phi) \ \ \mathrm{where} \ \ \mu = \int_0^\infty \phi  \frac{1}{\lambda^{5}} \frac{1}{e^{\frac{\Lambda}{\lambda}} -1} d\lambda
\end{equation}
and
\begin{equation}
\label{betaSimp}
\beta = \Lambda \frac{\kappa(\Lambda,\phi)}{\mu(\Lambda,\phi)}  \ \ \mathrm{where} \ \ \kappa = \int_0^\infty \phi  \frac{1}{\lambda^{6}} \frac{e^{\frac{\Lambda}{\lambda}}}{(e^{\frac{\Lambda}{\lambda}} -1)^2} d\lambda.
\end{equation}

Thus by making the (strong but widely relevant) approximation of a blackbody spectrum we have reduced the full behaviour of beaming signal down to two concise and well-behaved integrals which encapsulate all the information about the properties (namely temperature and radius) of the source and the filter used to observe it.

From here onwards we will retain the assumption that the spectrum is well represented by a blackbody.

\subsection{Defining the window function}

Depending on the application the optimal window function for an observation can vary widely. Furthermore physical constraints mean that the window function used in an observing instrument (whether it is a space telescope or a human eye) is never a perfect manifestation of that optimum.

I shall explore a few examples, chosen as much to be calculable as representative. Depending on the application, it is possible that none of the approximations given are accurate and for some cases we must fall back on a numerical expression for $\phi$ and a computational solution for equations \ref{fSimp} and \ref{betaSimp}.

\subsubsection{A $\delta$ function}
%maybe drop this section

If a filter is very sharply peaked it may resemble a $\delta$ function. Such a filter will be very sensitive to variations in the spectrum on scales of comparable width to itself and greater.

If
\begin{equation}
\phi(\lambda) = \delta(\lambda - \lambda_0)
\end{equation}
then
\begin{equation}
\mu_0 = \frac{1}{\lambda_0^5 (e^\frac{\Lambda}{\lambda_0} - 1)}
\end{equation}
and
\begin{equation}
\kappa_0 = \frac{e^\frac{\Lambda}{\lambda}}{\lambda_0^6 (e^\frac{\Lambda}{\lambda_0} - 1)^2}
\label{kappaSimp}
\end{equation}
where I have used the subscript $0$ to denote quantities evaluated at $\lambda=\lambda_0$

This gives
\begin{equation}
F_0=\frac{2\pi h c^2}{D^2}R^2 \frac{1}{\lambda_0^5 \Big(e^\frac{\Lambda}{\lambda_0} -1\Big)}
\end{equation}
and
\begin{equation}
\beta_0=\frac{\Lambda}{\lambda_0}\frac{e^\frac{\Lambda}{\lambda_0}}{e^\frac{\Lambda}{\lambda_0} - 1}
\label{beta0}
\end{equation}

These are the expressions derived in \citet{Loeb03} and are  approximately true for a filter with finite width also.

\subsubsection{Narrow box filter}

The next most tractable solution is a uniform filter centred around $\lambda_0$ with some width $\Delta \lambda$ (assumed $\ll \lambda_0$:
\begin{equation}
\phi(\lambda) = \begin{cases}
\frac{1}{\Delta \lambda} & \lambda_0-\frac{\Delta \lambda}{2}<\lambda<\lambda_0 + \frac{\Delta \lambda}{2}\\
    0& \mathrm{otherwise}
\end{cases}
\end{equation}
which reduces to the solution for a $\delta$ function as $\Delta \lambda \rightarrow 0$ (whilst maintaining $\int \phi d\lambda = 1$).

It will be useful to define $l(\lambda)=\frac{\Lambda}{\lambda}$ and $q(\lambda)=\frac{e^l}{e^l - 1}$ and I will subscript $0$ to denote the  function evaluated at $\lambda_0$.

We can find\footnote{Using a Taylor expansion and the fundamental theorem of calculus to express $\int_a^{a+\varepsilon} f(x) dx$ as $\varepsilon f(a) + \frac{\varepsilon^2}{2} \frac{df}{dx}(a) + \frac{\varepsilon^3}{6} \frac{d^2f}{dx^2}(a) + O(4)$ (where $\varepsilon \ll a$)}
\begin{equation}
\mu = \mu_0 \left(1 + \left(\frac{\Delta \lambda}{\lambda_0}\right)^2 
\left[ \frac{5}{4} - \frac{l_0 q_0}{2} + \frac{l_0^2 q_0^2}{24}\left(1+e^{-l_0}\right) \right] \right) + O(4)
%\mu = \frac{\Lambda}{\lambda_0} \frac{e^\frac{\Lambda}{\lambda_0}}{e^\frac{\Lambda}{\lambda_0}-1} \left(1 + \frac{\Delta \lambda}{2 \lambda_0}\left[ \frac{\Lambda}{\lambda_0} \frac{1}{e^\frac{\Lambda}{\lambda_0}-1} - 5\right] \right)
\end{equation}
(from which we can trivially find $F$)
and
\begin{equation}
\begin{aligned}
\kappa = \kappa_0 \Bigg(1 + \left(\frac{\Delta \lambda}{\lambda_0}\right)^2 
\bigg[ \frac{7}{4} -& \frac{7 l_0 q_0}{12}\left(1+e^{-l_0}\right) \\
&+ \frac{l_0^2}{24} \left(1+6 q_0^2 e^{-l_0}\right) \bigg] \Bigg) + O(4)
\label{kappaFull}
\end{aligned}
\end{equation}
giving
\begin{equation}
\begin{aligned}
\beta = \beta_0 \Bigg(1 + \left(\frac{\Delta \lambda}{\lambda_0}\right)^2 
\bigg[\frac{1}{2} -& \frac{l_0 q_0}{12}\left(1+7e^{-l_0}\right) \\
&+ \frac{l_0^2}{24} + \frac{l_0^2 q_0^2}{24} \left(5 e^{-l_0} -1\right)  \bigg] \Bigg) + O(4)
\end{aligned}
\label{beta1}
\end{equation}

More general filters may be analytically calculable, and any filter should be numerically well behaved.

In appendix \ref{approx} I discuss a small variation on the above model which can be used to find a more accurate directly calculable expression for a box-like filter with non-negligible width.

With this behaviour in hand we can move to applying this to orbiting systems.

\subsection{Beaming of the star}

Moving to our system of interest, imagine a star and a companion in an orbit, with semi-major axis $a$. The characteristic velocity of the star (assuming $M_c \ll M$) is
\begin{equation}
\label{velocity}
v \approx \frac{M_c}{M} \sqrt{\frac{G M}{a}}.
\end{equation}
For the purpose of this work it will be sufficient to use this as the magnitude of the observed line of sight velocity of the star, $v$. The full expression, including time-dependance, projection and eccentricity, is 
\begin{equation}
v(t) = \frac{M_c}{M} \sqrt{\frac{G (M+M_c)}{a (1-e^2)}} \sin{\theta_v} \left( \cos \left(\Phi(t) - \phi_v \right) - e \sin{\phi_v} \right).
\end{equation}
Here $e$ is the eccentricity and $(\theta_v,\phi_v)$ are the polar and azimuthal viewing angles respectively, in a frame where periapse of the companion is aligned with $(\frac{\pi}{2},0)$. $\Phi(t)$ is the time-varying phase angle of the companions orbit. Thus the signal amplitude depends on inclination through $\sin \theta_v$.

Thus relativistic beaming will cause a variation in the light curve of the star (ignoring the contribution of the companion) with characteristic magnitude
\begin{equation}
\delta \sim \beta \frac{M_c}{M} \sqrt{\frac{G M}{a c^2}}.
\end{equation}
The full signal will vary over an orbit, sinusoidally for a circular trajectory, and moving to a sudden bright peak for a very eccentric orbit (or a sudden flux dip if the object is moving away from the observer at the moment of pericenter passage). 

%As an example of the magnitude of such a signal, take a hot-Jupiter with $M_c \approx 0.01 M_\odot$ on a very close orbit ($a \approx 0.05 AU$) around a solar mass star ($M=M_\odot$). This gives $\delta \sim 20 ppm$, on the very edge of observability with modern space telescopes.

\subsection{Beaming of the companion}

The relativistic beaming signal in stars are already well documented and actively searched for \citep{Loeb03,vanKerkwijk10,Bloemen11,Lillo-Box16}. However, the crux of this paper is the possibility of observing beaming signals in the companion.%, and the information that can tell us about its temperature.

If we are comparing two objects which are spatially indistinguishable it is the overall variation, $\delta F(\lambda)$, that we must compare, where:
\begin{equation}
\delta F(\lambda) = \delta \cdot F = \frac{2 \pi h c}{D^2} \sqrt{\frac{GM}{a}} \frac{M_c}{M} R^2 \Lambda \kappa
\end{equation}
where $\Lambda$ and $\kappa$ are defined in equations \ref{Lambda} and \ref{betaSimp} respectively.

It is useful to define the ratio of the absolute flux variation from the companion compared to the star:
\begin{equation}
\label{ratio}
\left| \frac{\delta F_{c}}{\delta F} \right| = \frac{M}{M_c} \left( \frac{R_c}{R} \right)^2 \frac{\Lambda_c}{\Lambda} \frac{\kappa_c}{\kappa}
\end{equation}
where I have used the subscript $c$ for all properties relating to the companion. 

Splitting up the terms on the right hand side for a typical hot-Jupiter - the mass ratio can be very large ($M \lesssim 10^3 M_c$), the radius ratio small ($R^2 \gtrsim 10 R_c^2$) and the magnitude of the remaining term will depend on the temperature of the two objects and the window function.

Using the simplest approximation to $\kappa$ from equation \ref{kappaSimp} I find
\begin{equation}
\frac{\Lambda_c}{\Lambda}\frac{\kappa_c}{\kappa} = \frac{\Lambda_c}{\Lambda} e^\frac{\Lambda_c - \Lambda}{\lambda_0} \left( \frac{e^\frac{\Lambda}{\lambda_0} - 1}{e^\frac{\Lambda_c}{\lambda_0} - 1}\right)^2
\end{equation}
which captures all the temperature variation of the system, and varies smoothly as $\lambda_0$ changes.

This rightmost fraction is dependent on the temperatures and the wavelength in question, but can be evaluated at wavelengths smaller than $\Lambda$ or greater than $\Lambda_c$ (assuming here that $T > T_c$, though similar arguments can be made if this assumption is broken). For $\lambda \ll \Lambda$ (i.e. in the Wien tail of the primary)
\begin{equation}
\label{thetaLow}
\frac{\Lambda_c}{\Lambda}\frac{\kappa_c}{\kappa}\Big|_{\lambda_0} \rightarrow \frac{\Lambda_c}{\Lambda} e^{-\frac{\Lambda_c - \Lambda}{\lambda_0}} \rightarrow 0
\end{equation}
which tends to 0 at small wavelengths, i.e. the beaming signal of the star is completely dominant at small wavelengths.

At long wavelengths, $\lambda \gg \Lambda_c$  (i.e. in the Rayleigh Jeans tail of the companion),
\begin{equation}
\frac{\Lambda_c}{\Lambda}\frac{\kappa_c}{\kappa}\Big|_{\lambda_0} \rightarrow \frac{\Lambda}{\Lambda_c}  e^{\frac{\Lambda_c - \Lambda}{\lambda_0}} \rightarrow \frac{T_c}{T}
\end{equation}
and thus tends to a wavelength independent value which depends on the relative temperatures of the two bodies.

\begin{figure}
	\includegraphics[width=\columnwidth]{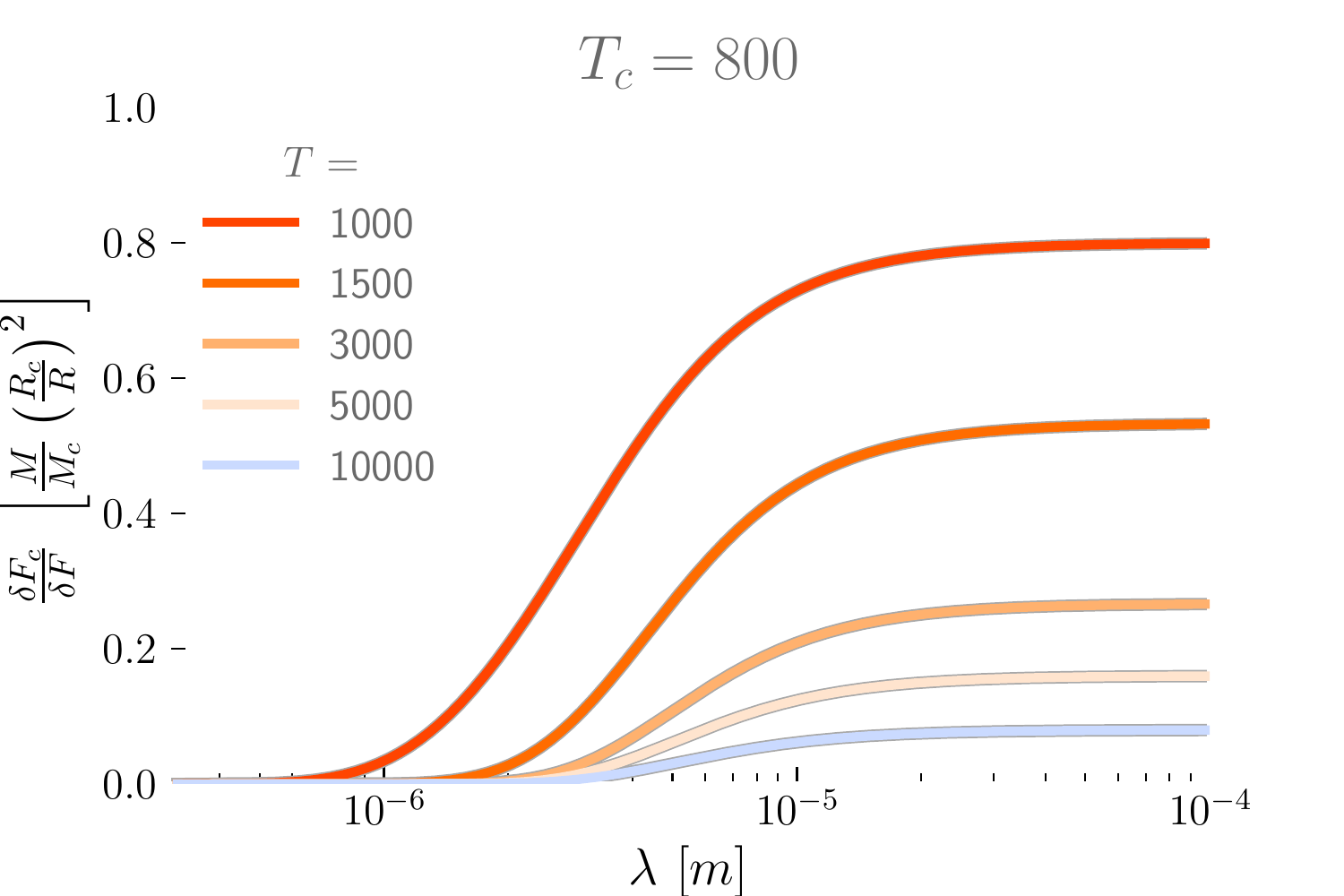}
    \caption{The ratio of characteristic beaming signal amplitude of a companion with $T_c=800 K$ compared to 4 stars of varying temperature. The ratio asymptotes to $\frac{T_c}{T}$ at high wavelengths. At lower wavelengths, it decreases.}
   	\label{frac800}
\end{figure}

In figure \ref{frac800} I show the behaviour of equation \ref{ratio} for a companion with $T_c = 800 K$. Thus for any system with a companion whose temperature is only a few times lower than the star's the companion may dominate the beaming signal at sufficiently long wavelengths.

\begin{figure}
	\includegraphics[width=\columnwidth]{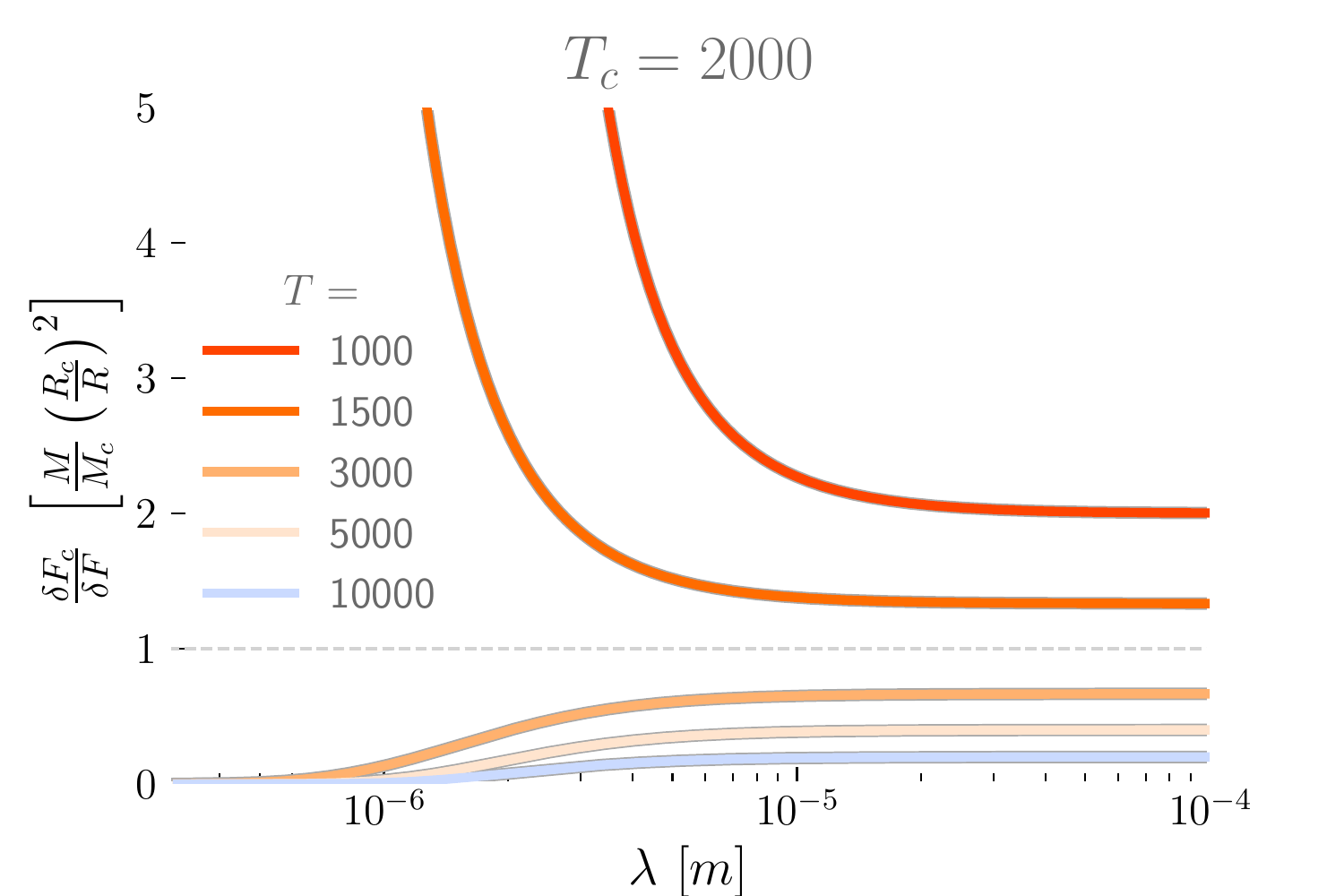}
    \caption{Similar to figure \ref{frac800} but with $T_c=2000 K$. Now the beaming signal from the companion is very large at low wavelengths if $T< T_c$ and approaches $\frac{T_c}{T}$ for large $\lambda$.}
   	\label{frac2000}
\end{figure}

Though I am mostly interested in this work in companions that are non-luminous, such as planets, we can also examine the relationship for hot objects. For comparison in figure \ref{frac2000} I show the same relationship with $T_c = 2000 K$, which might be appropriate for a companion heated by internal processes (for example a white-dwarf companion). Here some stars shown have $T<T_c$ and thus the companion dominates the ratio at all wavelengths, with the contribution decreasing at large wavelengths. For the rest of this work I will focus on companions with $T>T_c$.

The observable amplitude will be the sum of that of the signal of the companion and the star, however as the line of sight velocities of the two bodies are opposite the signals have opposing signs:
\begin{equation}
|\delta_\Sigma| = \left|\frac{\delta F - \delta F_c}{F + F_c}\right| =\frac{v}{c}\frac{\Lambda \kappa}{\mu} \left|\frac{1 - \frac{M}{M_c}\left(\frac{R_c}{R}\right)^2 \frac{\Lambda_c}{\Lambda}\frac{\kappa_c}{\kappa} }{1+\left(\frac{R_c}{R}\right)^2 \frac{\mu_c}{\mu}}\right|.
\end{equation}

\begin{figure}
	\includegraphics[width=1.0\columnwidth]{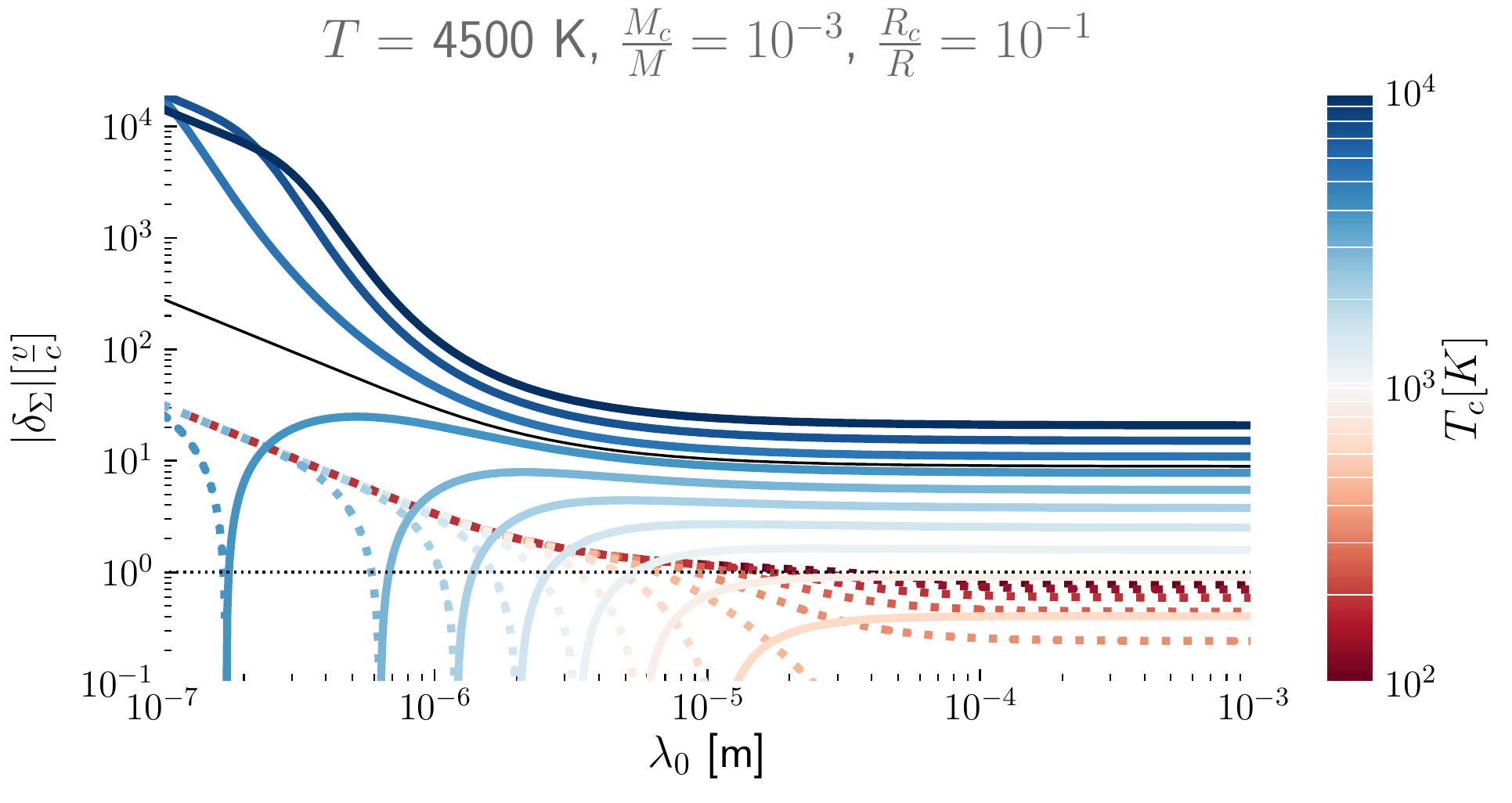}
    \caption{Variation in the total amplitude of the beaming signal with wavelength. For a given $T_c$ lines dashed denote where the signal is dominated by the star and solid lines where the planet gives the largest contribution. $\delta_\Sigma$ here is expressed in units of $\frac{v}{c}$, thus for a known stellar velocity we can rescale the dotted black line to find the observable change in brightness.}
   	\label{deltaSigma}
\end{figure}

Figure \ref{deltaSigma} shows this behaviour for a planet with the mass and radius of Jupiter. Note that $\delta_c$ can be expressed in units of $\frac{v}{c}$, allowing us to re-scale simply for different orbital situations. For reference, a stellar velocity of 300 $ms^{-1}$ is reasonable for a hot Jupiter observed in an edge-on system giving $\frac{v}{c} \sim$ 1 ppm.

\subsection{Derivable quantities from $\delta_\Sigma (\lambda_0)$}

To more precisely express what properties we can derive from multiple observations of beaming at different wavelengths we can examine $\delta_\Sigma$ at low and high wavelengths and where the signal goes to 0.

For large $\lambda_0$ (ignoring the small factor in the denominator) the signal tends to a constant:
\begin{equation}
\delta_\Sigma \rightarrow \frac{v}{c}\left| 1-\frac{M}{M_c}\left(\frac{R_c}{R}\right)^2 \frac{T_c}{T}\right|.
\end{equation}

%\begin{equation}
%\delta_\Sigma \rightarrow \frac{v}{c} \left|\frac{1 - %\frac{M}{M_c}\left(\frac{R_c}{R}\right)^2 %\frac{\Lambda}{\Lambda_c}}{1+\left(\frac{R_c}{R}\right)^2 %\frac{\Lambda}{\Lambda_c}}\right|
%\end{equation}
%which tends to a constant.

%We can approximate this one step further,

As $\lambda_0 \rightarrow 0$ I find 
\begin{equation}
\delta_\Sigma \rightarrow \frac{v}{c} \frac{h c}{\lambda_0 k T}.
\end{equation}
%\begin{equation}
%\delta_\Sigma \rightarrow \frac{v}{c} \left|\frac{1 - %\frac{M}{M_c}\left(\frac{R_c}{R}\right)^2 %\frac{\Lambda_c}{\Lambda}e^{-\frac{\Delta \Lambda}{\lambda_0}}}{1+\left(\frac{R_c}{R}\right)^2 %e^{-\frac{\Delta \Lambda}{\lambda_0}}} \right|
%\end{equation}

And $\delta_\Sigma = 0$ when
\begin{equation}
\frac{T_c M_c}{R_c^2} \sinh\left( \frac{h c}{2\lambda_0 k T_c}\right) = \frac{T M}{R^2} \sinh\left( \frac{h c}{2\lambda_0 k T}\right)
\end{equation}
(when this has no real solutions we have a system for which the star always dominates).

Thus a precise observation of the beaming signal at multiple wavelengths is enough to tell us $\frac{v}{c}$, $T_c$ and $\frac{M}{M_c}\left(\frac{R_c}{R}\right)^2$ (assuming the stellar temperature is known to reasonable accuracy beforehand). The exception to this is the case where the beaming signal of the companion is always negligible, then we can only determine $\frac{v}{c}$.

If any two of the stellar mass, semi-major axis or period are well constrained by other methods the third can be inferred and the value of $\frac{v}{c}$ gives us a specific value of $M_c$ (see equation \ref{velocity}). Thus the observable parameters which we can derive from an arbitrarily accurate observation become $M_c \sin \theta_v$, $T_c$ and $\left(\frac{R_c}{R}\right)^2$. Similarly if  $\left(\frac{R_c}{R}\right)$ is well constrained (e.g. from a transit) we can infer $v$, $T_c$ and $\frac{M_c}{M}$.

\begin{figure}
	\includegraphics[width=1.0\columnwidth]{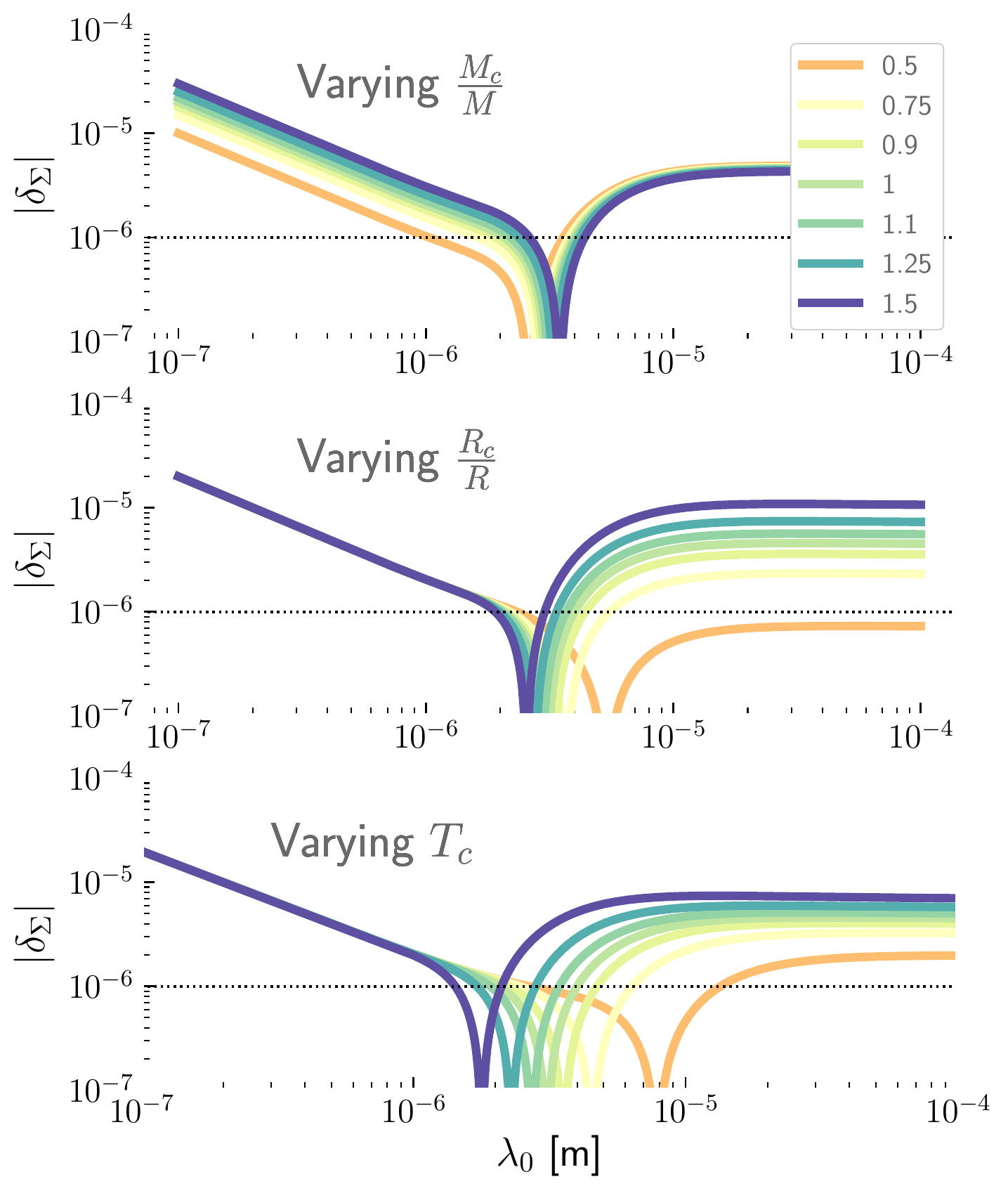}
    \caption{Using NGTS-1b (\citealt{Bayliss18}, a hot-Jupiter on a 2.6 day orbit around a K-type star) as an example system I explore how the beaming signal changes as I vary the system properties. Each line shows the initial system varied by a factor ranging from 0.5 - 1.5, as shown in the legend of the upper plot. It can be seen that the amplitude of the beaming signal from the star is sensitive to the mass ratio, the amplitude of the companions signal depends strongly on the radius ratio and the wavelength at which the signal cancels (as well as the companion beaming amplitude) depend strongly on the companions temperature.}
   	\label{variation}
\end{figure}

Figure \ref{variation} show explicitly, using an example confirmed exoplanet - NGTS-1b \citealt{Bayliss18} - the variation of $\delta_\Sigma$ with wavelength and how this profile changes as I vary the parameters of the system.

For an otherwise undetermined system the wavelengths of interest are not known a priori, and observations must be taken over a number of wavelengths to find where $\delta_\sigma \rightarrow 0$.

These signals are small, in most cases, but universal. Any orbiting bodies will display them, though possibly at such low amplitude that they will be drowned out in the noise. Thus the question of interest is whether detecting these signals is possible with current and future instruments, and what such a detection could reveal.

%\begin{figure*}
%	\includegraphics[width=1.0\textwidth]{plots/delta.pdf}
%    \caption{The magnitude of relativistic beaming signals that would be observed in the population of known exoplanets if they had a temperature of 800 K (upper) or 2000 K (lower) and are observed at 3 $\mu m$ (left) or $10 \mu m$ (right) wavelengths. Each point represents a known planet, the colour shows the temperature of the star (see figure \ref{frac800} for reference) and the size represents the relative radius of the planet and star (larger points, larger planets). The dashed line shows when the amplitudes would be equal (and thus the signal will cancel).}
%   	\label{deltaFig}
%\end{figure*}

\section{Observational possibilities}
\label{observation}

There are many systems we could observe or imagine where relativistic beaming could be significant, such as binary star systems \citep{vanKerkwijk10}, brown dwarves \citep{Lillo-Box16} or compact objects \citep{Bloemen11} and is theorized as being a plausible method for detecting stellar mass black holes in binaries \citep{Masuda18}. For the purpose of discussion in this work I will focus on a specific sample as a case study - the current known population of exoplanets.

For most planets it will be a reasonable assumption that their surface temperature is set by the temperature of their star and the relative distance, often called the equilibrium temperature:
\begin{equation}
T_{c,eq} = T \sqrt{\frac{R_c}{2 a}}
\label{Teq}
\end{equation}
(in its simplest form, ignoring albedo).

%However, when a young astronomical body forms via gravitational collapse there is also significant kinetic energy converted to heat. Compared to typical lifetimes of stars and planets this excess temperature is radiated away on short timescales. However depending on how much energy is injected and how the body cools (two complex and multifaceted questions which we do nothing but skim past here) we may be able to identify hot, newly formed planets.

To get a view (though far from a full one) of the range of possible signals from such hot young planets I will use the properties of known planets: the confirmed exoplanets according to the NASA exoplanet archive. Specifically, I use planets with known $M, M_c, R, R_c$ and $T$ and an assumption about $T_c$ to estimate the beaming signal of both the star and the companion. %For comparison we consider two possible values of $T_c$: $800 K$ (the suggested temperature of some known directly imaged young planets) and $2000 K$.

\begin{figure*}
	\includegraphics[width=1.0\textwidth]{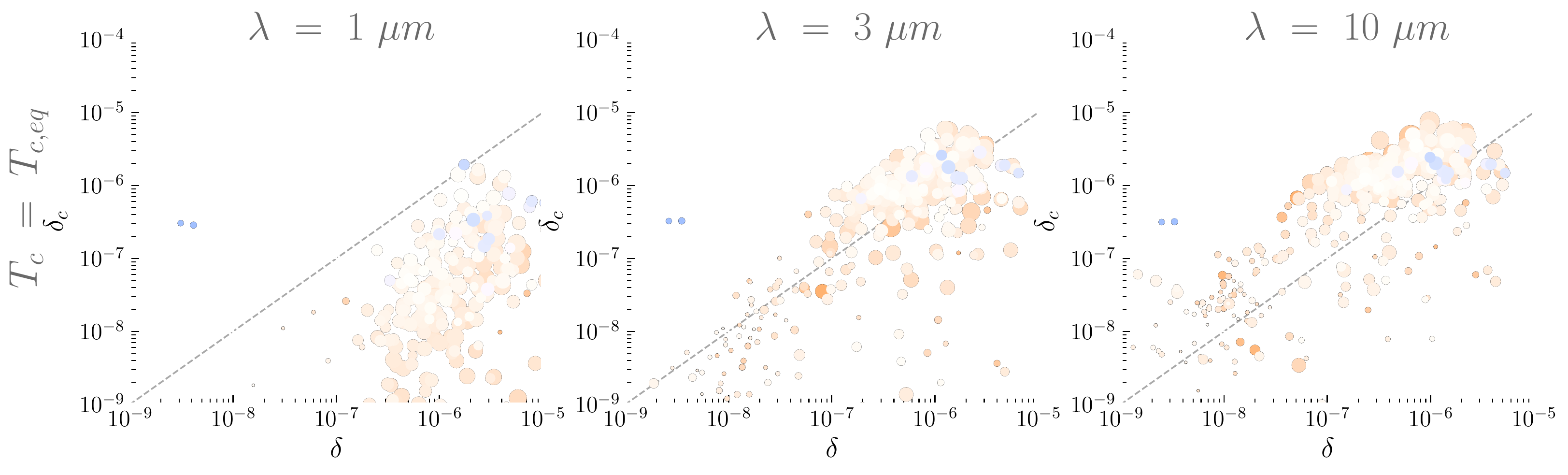}
    \caption{The magnitude of relativistic beaming signals that would be observed in the population of known exoplanets if their equilibrium temperature follows equation \ref{Teq}. Each point represents a known planet, the colour shows the temperature of the star (see figure \ref{frac800} for the colour legend) and the size represents the relative radius of the planet and star (larger points, larger planets). The dashed line shows when the amplitudes would be equal (and thus the signal will cancel). At $1 \mu m$ (the upper wavelength of a survey like TESS) the star is still completely dominant, but at only slightly higher $\lambda$ signal from the companion can become visible.}
   	\label{deltaEq}
\end{figure*}

Figure \ref{deltaEq} shows the characteristic beaming amplitude we would observe for all these systems, if the planet is at its predicted equilibrium temperature. It can be seen that for current exoplanet surveys (e.g. TESS which spans from 600 to 1000 nm) the star will dominate. However, we only need go to slightly higher wavelengths and the companion dominates, at a level that we might expect to be observable relatively soon.

%Figure \ref{deltaFig} shows the characteristic beaming amplitude we would observe for all these systems, at two different wavelengths in the infra-red (for reference the characteristic wavelengths of the $T_c = 800$ and $2000 K$ companions are $\Lambda \approx 18$ and $7 \mu m$ respectively).

The variations are small, at largest of order $\sim 10$s of ppm (parts per million) but signals of this magnitude can be detectable, increasingly so with newer surveys. As we would expect, hotter companions observed at longer wavelengths exhibit larger signals (though moving to yet larger wavelengths will have little effect, due to the asymptotic nature of equation \ref{ratio}).

The current and next generation of space telescopes is capable of resolving changes in local stellar brightness of order of 100's, sometimes even 10's, of ppm. For example the IRAC detector on the Spitzer mission has been used to make observations with a precision of 30 ppm \citep{Beichman14}. Future instruments such as JWST (estimated at 10 ppm for a 1 hour integration, \citealt{Beichman14}) and TESS (60 ppm for 1 hour observing nearby stars, \citealt{Ricker15}) will give yet greater precision, allowing us to look for these signals in more and more systems. Sources of astrophysical and systematic noise will limit the achievable precision, especially for longer period planets where the lightcurve is effectively more sparsely sampled. Robust understanding and modelling of this noise is a necessary, non-trivial challenge that is being discussed throughout the exoplanet and stellar observation communities (see for example \citealt{Aigrain17,Jansen18,Prsa19,Hippke19,Wheeler19} and the references therein).

Figure \ref{variation} shows, using NGTS-1b  as an example system, the sensitivity of the beaming signal to changes in the system parameters. It can be seen that each parameter transforms the profile in a unique way, and thus for sufficiently precise measurements at a range of infra-red wavelengths, we could theoretically constrain each without degeneracies (given that it's a known period transiting planet).

\begin{figure}
	\includegraphics[width=1.0\columnwidth]{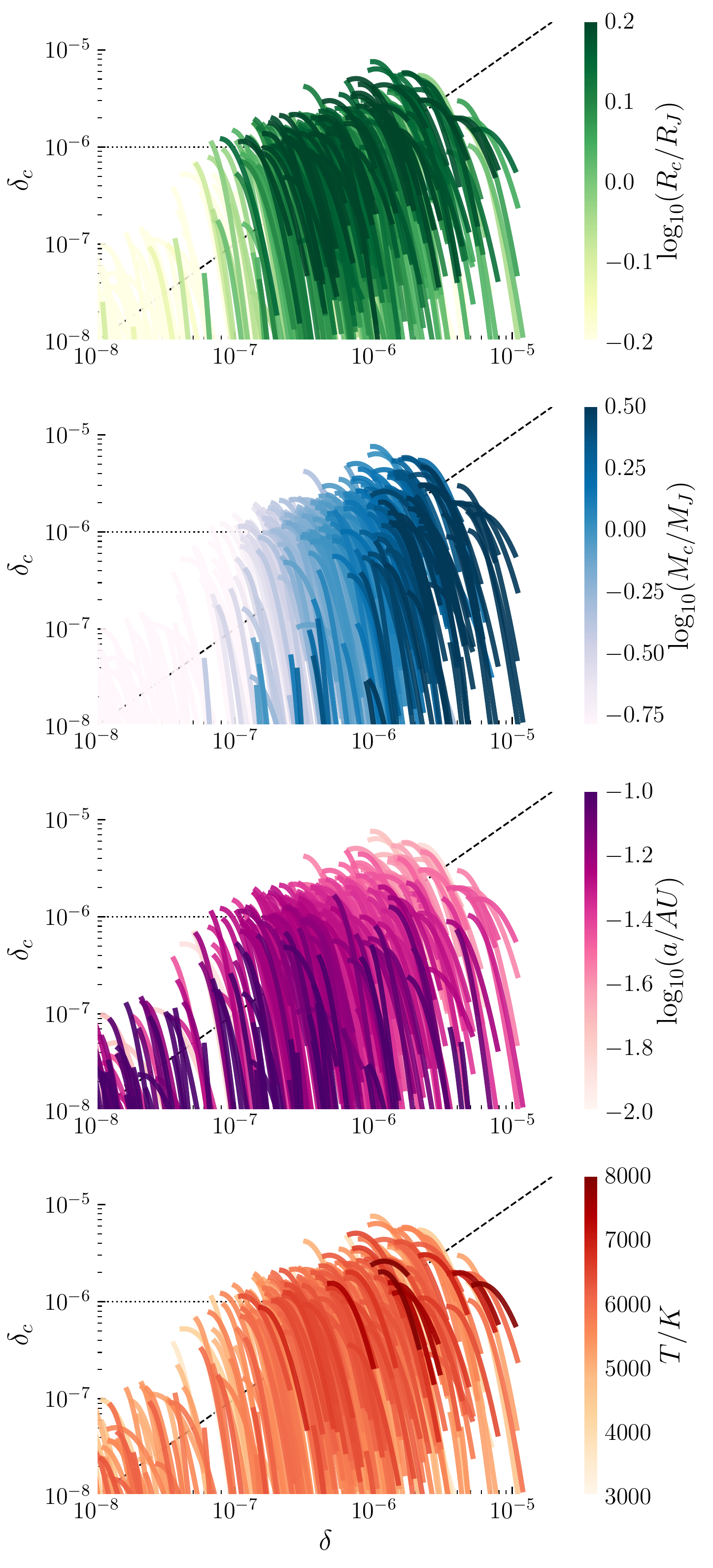}
    \caption{For every confirmed exoplanet I compare the beaming signal from the host star ($\delta$) and its companion ($\delta_c$) over wavelengths from $1 \mu$m to $10 \mu$m - which traces a curved line for which the primary is most dominant at low wavelengths. Each system is coloured by the properties of the star and planet to facilitate a qualitative intuition about which properties lead to large beaming signals for the star and companion. Along the dashed line the signals are equal (and hence cancel) and the dotted line shows 1 ppm amplitude.}
   	\label{properties}
\end{figure}

\begin{figure}
	\includegraphics[width=1.0\columnwidth]{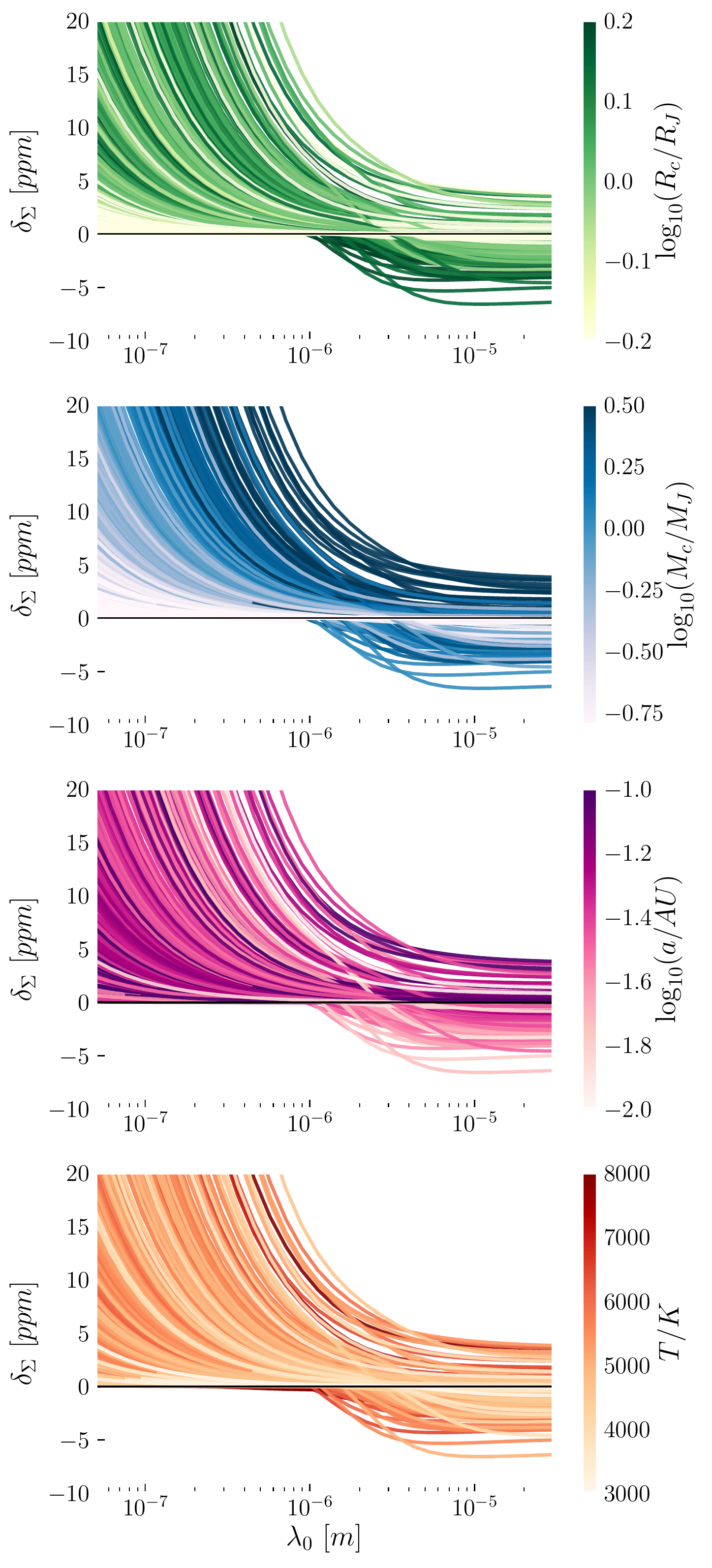}
    \caption{Similar to figure \ref{properties} but now showing the variation of the total beaming signal, $\delta_\Sigma$ (positive when the star dominates, negative for the companion), as a function of observed wavelength.}
   	\label{propertiesLambda}
\end{figure}

Figures \ref{properties} and \ref{propertiesLambda} explore how the magnitude of the beaming effect varies with planetary properties. It can be seen that the beaming signal is strongly dependant on planetary radius and semi-major axis, and thus will be most observable for hot-Jupiters. The mass of the companion has a greater effect on the beaming signal for the star, with the most massive systems having large, stellar dominated signals at all wavelengths. There is no discernible structure in the plots against stellar temperature, as the companions equilibrium temperature is also affected by increasing stellar temperature and the two effects will approximately cancel.

%It will be a fascinating insight into planetary formation whether we convincingly find such signals, or can rule them out. The place of Hot-Jupiters in the evolutionary history of a planetary system is far from known, and being able to measure (or put an upper limit on) their temperature will constrain their formation channels considerably.

The sample of systems chosen here for this case study is relatively conservative, especially as I have ignored effects such as eccentricity (which can only increase the amplitude). More extreme systems, or even completely different pairs of objects (such as those including brown dwarfs or compact objects) may be prime targets for a similar exploration.% We leave it up to each and any reader to consider the relevance of these signals for any astrophysical orbit of interest.

\subsection{Complications}

Finally, it is worth noting that for many of the planets shown here, and over much of the possible parameter space for a star and a companion, the beaming signal will not be the dominant source of variation in the light curve. Most systems shown here are transiting planets, a much larger effect, but as shown in \citet{Penoyre18b} tidal effects and reflections of the stars light by the planet are dominant sources of phase variation over much of parameter space. 

Each out of transit effect peaks at different points of the orbit - reflections peak at opposition, tides vary twice over one orbit and are minimal at conjunction and opposition, and beaming peaks at first (third) quadrature when the star (planet) dominates \citep{Shporer17}. Thus we may be able to extract the size of the beaming signal even when it is subdominant - though complications like orbital eccentricity and shifted planetary hotspots will lead to contamination.

If secondary eclipses are observed that provides another window into determining the temperature, and in most cases where both companion beaming and eclipses are seen the latter will be a larger and clearer signal. However, the fraction of the total population of planets that transits is very small - and the number of systems in which out-of-transit effects are visible may come to rival transiting systems as the precision of our telescopes increases.

In systems which do not transit we may also observe the modulation of the thermal emission of light due to the difference in temperature between the day and night side of the planet which in some cases, such as tidally locked planets, can be of similar magnitude to the equilibrium temperature. These emission signals give us a clear measure of the temperature difference, but not its absolute value. At long wavelengths the ratio of the strength of the beaming signal to the emitted light signal is at most $\frac{v}{c} \frac{M}{M_c} \frac{T_c}{\Delta T_c}$ (where $\Delta T_c$ is difference between day and night side temperature) - thus in many cases we would expect emission to dominate but each can tell us different information about the system.

It is also possible that, if the beaming signal can be well resolved, we may begin to see substructure as a function of observing wavelength, $\lambda_0$. This would correspond to changes in the effective temperature at particular wavelengths due to the atmospheric properties of the companion. Thus we may be able to probe the chemical composition and thermal profile of non-transiting planets, and better understand their atmospheres \citep{Gandhi19}.

Systems with significant beaming signals will likely be detected and characterized by other methods, but comparing the beaming signals will give strong constraints on the temperature of the companion and in some cases may be the best way of directly observing its presence.

\section{Conclusions}

In this paper I explore the photometric effect of relativistic beaming of a star, and for the first time include the effect on the companion also.

For a star and a companion with two different temperatures each may dominate the beaming signal at different wavelengths. Assuming the star is hotter it dominates at low wavelengths, due to a much higher flux, and the companion can dominate at long wavelengths due to its higher velocity.

As the two objects are moving in opposite directions the signals can cancel, and then change sign, when observed at higher wavelengths. If we can measure the amplitude of the signal at low wavelengths (where the Wien tail of the primary dominates) and at high wavelengths (where the Rayleigh Jeans tail of the companion can dominate), along with the wavelength at which the signal cancels totally, we could uniquely constrain the orbital line of sight velocity of the primary, the temperature of the secondary and $\frac{M}{M_c}\left(\frac{R_c}{R}\right)^2$.

I present a representative case study, looking at the beaming signals we would expect for the population of known exoplanets. It is feasible that these signals could be detectable with the next generation of space telescopes, such as JWST, giving new constraint and insight into the properties of extrasolar planets.

\section*{Acknowledgements}

Z.P. acknowldeges support from the UK Science and Technologies Facilities Council (STFC) and would also like to thank Cathie Clarke, Nikku Madhusudhan, Simon Hodgkins and Emily Sandford for their suggestions and feedback.

%%%%%%%%%%%%%%%%%%%%%%%%%%%%%%%%%%%%%%%%%%%%%%%%%%



\bibliographystyle{mnras}
\bibliography{bib} % if your bibtex file is called example.bib

%\begin{thebibliography}{99}
%\end{thebibliography}

%%%%%%%%%%%%%%%%%%%%%%%%%%%%%%%%%%%%%%%%%%%%%%%%%%

%%%%%%%%%%%%%%%%% APPENDICES %%%%%%%%%%%%%%%%%%%%%

\appendix

\section{Accurate approximations to $\beta$}
\label{approx}

Here I briefly discuss the accuracy of approximate solutions to equation \ref{betaSimp}. Though integrating numerically is simple enough, in many cases this introduces an extra computational cost that we may wish to avoid.

I have shown two approximate versions of $\beta$, accurate to zeroth order (equation \ref{beta0}) and second order (equation \ref{beta1}) in $\frac{\Delta \lambda}{\lambda_0}$.

\begin{figure}
	\includegraphics[width=\columnwidth]{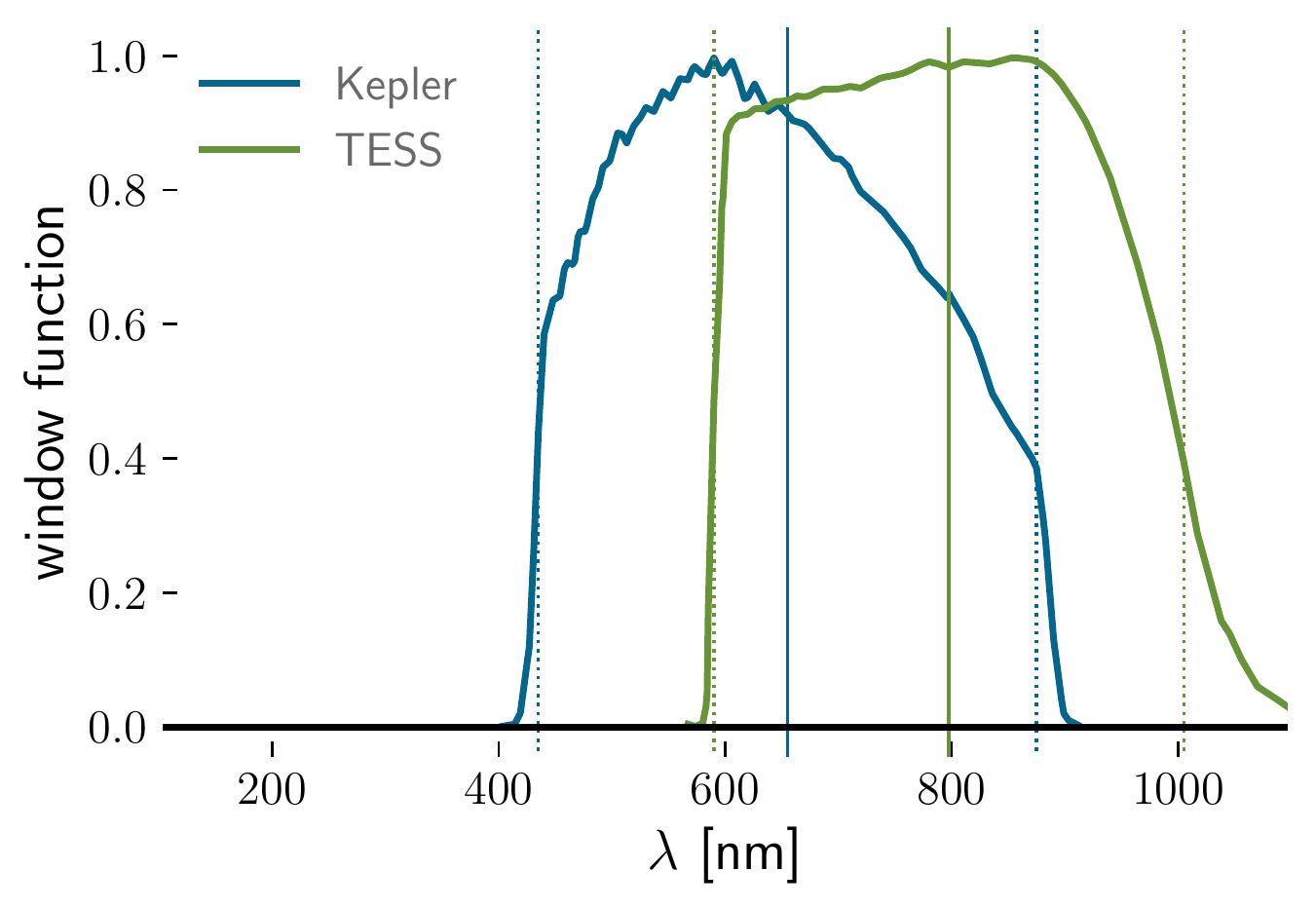}
    \caption{The window functions (normalised to a maximum value of 1) for the Kepler and TESS surveys. Dotted vertical lines show the wavelengths at which the response falls below $\frac{1}{e}$, giving an approximate $\Delta \lambda$ (the difference) and $\lambda_0$ (the midpoint, solid vertical line).}
   	\label{windowFunctions}
\end{figure}

Let us compare with the results we would derive for a numeric integration of equation \ref{betaSimp} for two relevant filters, those of the Kepler and TESS surveys (shown in figure \ref{windowFunctions}). Both are relatively wide and neither are very well represented by a box-car function. Nethertheless I can derive reasonable approximations to $\beta$ that well fit the numerical evaluation using these filters.

\begin{figure}
	\includegraphics[width=\columnwidth]{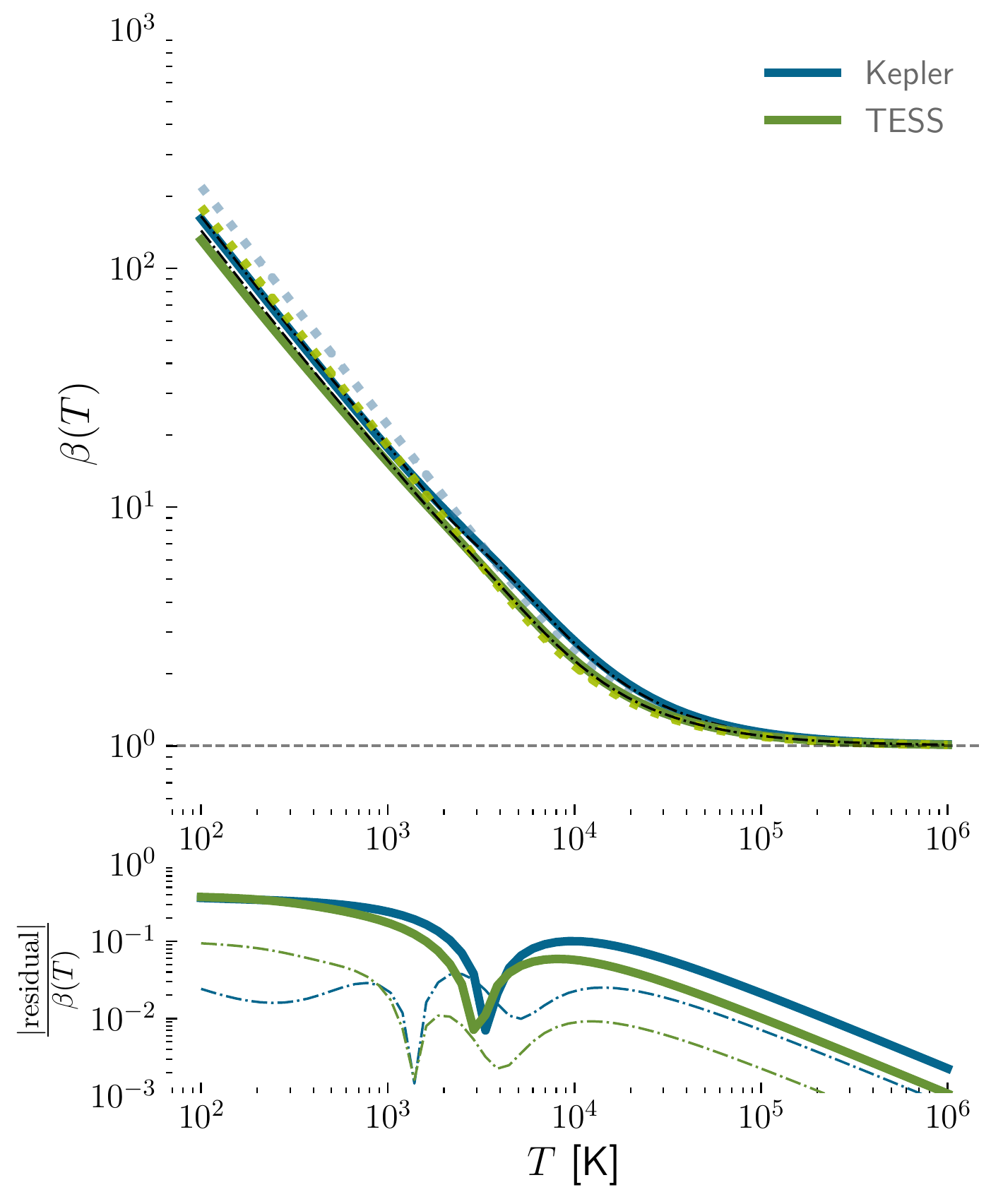}
    \caption{Top panel -  the value of $\beta$ derived for blackbodies of varying effective temperature, derived by numerical integration (solid lines), zeroth order approximation (dotted colour) and second order approximation with $\sigma$ correction (dashed black).
    Bottom panel - the relative error, compared to the integrated value, due to approximating to zeroth order (solid line) and second order (dashed).}
   	\label{betaTemp}
\end{figure}

Figure \ref{betaTemp} shows the value of $\beta$ derived by these approximations and the relative error. The zeroth order approximation is relatively accurate at high temperatures, but produces errors $ > 10 \%$ below $10,000$ K.

The second order approximation (not shown) performs better but still breaks down at the lowest temperatures, producing errors $>10\%$ below $1000$ K.

The reason that these approximations break down at low $T$ (and thus large characteristic wavelengths, $\Lambda$) is that below the peak in $F_\lambda$ the integral is dominated by the high wavelength end of the filter. Thus I suggest a (very approximate) correction, to give more accurate results at low temperatures without significantly increasing error at high temperatures.

At low $T$ (and high $\Lambda$) the upper end of the interval dominates the behaviour. Thus I suggest replacing $\lambda_0$ and $\Delta \lambda$ in equation \ref{beta1} with
\begin{equation}
\lambda_0' = \lambda_0 + \sigma(\Lambda) \frac{\Delta \lambda}{2}
\end{equation}
and
\begin{equation}
\Delta \lambda' = \Delta \lambda (1 - \sigma(\Lambda))
\end{equation}
where $\sigma$ is a sigmoid-like function which goes to 0 for high temperatures and 1 at low temperatures.

It may be possible to derive an accurate approximation to $\sigma$ based on the characteristics of the system, but for the moment I will simply use a function picked by hand to approximate the behaviour:
\begin{equation}
\sigma = \frac{1}{1-e^{-5(\log_{10}(\frac{\Lambda}{30}) - \log_{10}(\lambda_0))}}.
\end{equation}

In the lower panel of figure \ref{betaTemp} I show that this approximation keeps the error below $\sim 10\%$ for all temperatures.

%%%%%%%%%%%%%%%%%%%%%%%%%%%%%%%%%%%%%%%%%%%%%%%%%%

% Don't change these lines
\bsp	% typesetting comment
\label{lastpage}
\end{document}